\documentclass[pre,aps,showpacs,nofootinbib,twocolumn]{revtex4}
\usepackage{amssymb}
\usepackage{graphicx}

\input{tcilatex}

\begin{document}

\title{ $\pi$-Josephson Junction and Spontaneous Superflow \\
in Rings from Ultracold Fermionic Atomic Gases}
\author{Miodrag L. Kuli\'{c}}

\address{J. W. Goethe-Universit\"{a}t Frankfurt am Main, Theoretische Physik,
Max-von-Laue-Strasse 1, 60438 Frankfurt am Main, Germany}

\begin{abstract}
The BCS-like pairing in ultracold fermionic atomic ($UCFAG$) gases is
studied in the model of "isotopic-spin" pairing proposed in 1991 \cite%
{Ku-Hof-SSC}. This model assumes a mismatch ($\delta $) in chemical
potentials of pairing fermionic atoms. It is shown that a $\pi $-Josephson
junction can be realized in $UCFAG$ systems, where the left and right banks $%
S$ are the $UCFAG$ superfluids. The weak link $M$ consists from the normal $%
UCFAG$ with the finite mismatch $\delta $. If the $\pi $-junction is a part
of a closed ring the superfluid mass-current flows spontaneously in the
ring, i.e., the time-reversal symmetry is broken spontaneously. This is
realized if the radius of the ring $R$ is larger than the critical one $%
R_{c} $. All these effects exist also in the case when $\delta \gg \Delta $,
where $\Delta $ is the superfluid gap, but with the reduced thickness of the
weak link.

It is also discussed, that if junctions $SM_{1}M_{2}S$ and trilayers $%
M_{1}SM_{2}$ from $UCFAG$ are realizable this renders a possibility for a
novel electronics - \textit{hypertronics}.
\end{abstract}

\date{\today}
\maketitle

\section{\ Introduction}

Physics of ultracold atoms is fascinating in many respects. The ways how
these systems can be manipulated in magnetic and optical traps open
possibilities to an enormous number of physical effects which can be
realized and studied in these systems \cite{cold atoms-review}, \cite%
{Hofstetter}. Thanks to a large variety of experimental techniques one can
manipulate atomic gases by magnetic and electromagnetic fields giving rise
to realization of various macroscopic quantum effects. First of all, a
number of effects known in standard superfluids and superconductors were
already realized in ultracold atomic gases \cite{cold atoms-review}, \cite%
{Hofstetter}. In that sense a number of already developed theoretical
methods and ideas in solid state physics were used in physics of ultracold
atoms. On the other hand this fascinating field gives us new possibilities
in studying many aspects of the Bose-Einstein condensation (BEC) and of the
superfluidity of Fermi condensates (BCS) which are difficult to realize in
solid state physics. For instance in an ultracold gas of bosonic atoms, such
as $^{87}Rb$ or $^{7}Li$, one can study Bose-Einstein condensation(BEC). If
instead we deal with a gas of \textit{fermionic atoms}, such as $^{40}K$ and
$^{6}Li$, one can study not only fermionic superfluidity (BCS) but also the
transition BEC-BCS by directly controlling the interaction (scattering
length) between atoms \cite{Hofstetter}. Then by tuning magnetic field in
magnetic traps and/or electromagnetic field in optical lattices one can vary
the atomic scattering length $a_{F}$ in a broad range (especially for
energies just near the Feshbach resonance), that even BCS-like pairing of
ultracold atoms can be realized with $a_{F}<0$. It seems that this
possibility is already realized with ultracold fermionic alkali gases $%
^{40}K $ and $^{6}Li$ \cite{Greiner-12-Nature}, \cite{Jochim-13-Science}.
The cooling of \ a magnetically trapped spin-polarized $^{6}Li$ Fermi gas up
to $7\times 10^{7}$ atoms at $T<0.5$ $T_{F}$ and up to $3\times 10^{7}$
atoms at $T\approx 0.05$ $T_{F}$ is already realized. The strength of the
atomic interaction was controlled by applying a magnetic field and tuning to
Feshbach resonances, which occur when the total energy of interacting
particles in open channels is the same as the energy of a bound molecular
state in closed channels.

For such a two-level fermionic system it was proposed by the present author
and his collaborator in the paper "Inhomogeneous Superconducting Phase in
Absence of Paramagnetic Effect" a BCS-like model for the \textit{"isotopic
spin" pairing} of two (or more) species \cite{Ku-Hof-SSC} - the $ISP$ model.
The basic ingredient of the $ISP$ model is that two species (atoms,
electrons, quarks, nucleons, etc.)\ may have either different kinetic
energies, or different energy levels (or chemical potentials), i.e there is
a mismatch of energy levels of fermionic atoms which participate in pairing.
It turns out that this model is an adequate theoretical framework for a
number of experimental situations which deal with ultracold fermionic alkali
gases, such as for instance $^{40}K$ and $^{6}Li$ . The present experimental
techniques allow realizations of systems with two (or more) hyperfine
levels. They can be furthermore manipulated in order to maintain various
kinds of atomic pairing from BCS-like to strong coupling case as well as to
realize the mismatch between the paired "isotopic" atomic levels (chemical
potential). For further application of this model see below.

In the following we apply this model in studying nonuniform superfluidity in
ultracold fermionic atomic gases ($UCFAG$) with oscillating order parameter $%
\Delta (\mathbf{r})$ - which is due to the mismatch effect of two hyperfine
states of fermionic atoms. This effect is analogous to the
Larkin-Ovchinikov-Fulde-Ferrell ($LOFF$ or $FFLO$) phase in superconductors
\cite{LO-paper}, \cite{FF-paper}. Based on this effect we propose a novel
Josephson junction ($SMS$) where the left and right superfluid ($S$)\ gases
have uniform order parameters $\Delta _{L,R}=const$, while the weak link ($M$%
)\ with the mismatch effect is in the normal state. This gives rise to an
oscillating superfluid amplitude inside the weak link $M$ and as a result
the so-called $\pi $-junction can be realized. We show that if such a
junction is a part of the closed ring then \textit{spontaneous and
dissipationless superfluid current} can flow through the ring depending on
the size of the ring. In that case there is \textit{spontaneous breaking of
the time-reversal} in the system. Finally, we discuss possible realizations
and generalizations of this novel effect in ultracold fermionic gases.

\section{Model of isotopic-spin pairing \ and ultracold fermionic gases}

The model with the "isotope-spin" pairing ($ISP$) \cite{Ku-Hof-SSC} is a
generalization of the BCS pairing mechanism to systems with internal degrees
of freedom such as for instance, \textit{nuclear matter} - with isospin
numbers, \textit{quark matter} - with the color and flavor quantum numbers,
\textit{ultracold fermionic gases} - with quantum numbers of the hyperfine
atomic states, \textit{layered} and \textit{multiband superconductors} -
with quantum numbers enumerating layers and bands. The pairing constituents
(electrons, nucleons, quarks, neutral atoms, etc.) can be either charged or
not and in those cases we deal either with real superconductivity or with
superfluidity of matter. All these systems posses a natural mismatch in
energies (or masses) of constituents participating in pairing. This is an
important property since in the case when the mismatch parameter $\delta $
is of the order of the bare superconducting gap $\Delta _{0}$ a nonuniform
superfluidity (superconductivity) is realized - an analogue of the $LOFF$
state in metallic superconductors placed in the Zeeman field. The
Hamiltonian of the $ISP$ model is given by \cite{Ku-Hof-SSC}

\begin{equation}
\hat{H}=\hat{H}_{0}+\hat{H}_{BCS}  \label{ham}
\end{equation}%
\begin{equation}
\hat{H}_{0}=\sum_{a=1,\sigma }^{n}\int d^{d}x\hat{\psi}_{a\sigma }^{\dag }(%
\mathbf{x})\varepsilon _{a\sigma }(\mathbf{\hat{p}})\hat{\psi}_{a\sigma }(%
\mathbf{x})  \label{H0}
\end{equation}

\begin{equation}
\hat{H}_{BCS}=-\sum_{a,b}g_{ab}\int d^{d}x\hat{\psi}_{a\uparrow }^{\dag }(%
\mathbf{x})\hat{\psi}_{b\downarrow }^{\dag }(\mathbf{x})\hat{\psi}%
_{b\downarrow }(\mathbf{x})\hat{\psi}_{a\uparrow }(\mathbf{x})  \label{BCS}
\end{equation}%
The $ISP$ order parameter $\Delta _{ab}(\mathbf{x})$

\begin{equation}
\Delta _{ab}(\mathbf{x})=g_{ab}F_{ab}(\mathbf{x})=-g_{ab}\langle \hat{\psi}%
_{a\uparrow }(\mathbf{x})\hat{\psi}_{b\downarrow }(\mathbf{x})\rangle .
\label{gap}
\end{equation}

In the following we shall study the BCS-like pairing and Josephson effect in
ultracold fermionic alkaline gases ($UCFAG$)\ in which case $a,b=1,2$ and $%
\Delta _{ab}\neq 0$ for $a\neq b$ only - see the explanation below, while
the electronic spin states $\uparrow $ and $\downarrow $ do not enter (they
are accounted for via $a,b$). For simplicity we omit indices $a,b$ in $%
\Delta _{ab}(\mathbf{x})$, i.e. $\Delta _{ab}(\mathbf{x})\equiv \Delta (%
\mathbf{x})$. In the case of two ($a=1,2$)\ "isotopic" bands with the
mismatch in their energies (or chemical potentials) - the \textit{"isotopic"
band splitting }($\delta $) - the quasiparticle spectrum is given by
\begin{equation}
\varepsilon _{a=1,2}(\mathbf{\hat{p}})=\varepsilon (\mathbf{\hat{p}})\mp
\delta .  \label{mismatch}
\end{equation}%
Note, that the splitting parameter $\delta $ can in principle depend on
momenta too, but the results obtained below are qualitatively valid also in
this case.

It is interesting to mention that the above model and some physical results
\cite{Ku-Hof-SSC} were frequently used later on in studying diverse physical
problems such as: ($1$) \textit{ultracold} \textit{fermionic atomic gases} -
where $a,b=1,2,..n$ enumerate hyperfine atomic states of alkali atoms; ($2$)
in \textit{quantum chromodynamics} - where the problem of pairing in the
quark matter, i.e., the \textit{color superconductivity }($CS$), was studied
\cite{Rischke}. In the latter case the $CS$ pairing of the colored quarks is
mathematically more complicated than the standard BCS case, since the former
system is characterized by the matrix coupling constant $g_{ab}$ in the
internal degrees of freedom, where $a,b=(c,f,\sigma )$ are quantum numbers
of quarks: $c=1,2,3$ - \textit{color}; $f=1,2...n$ - \textit{flavor} (up,
down, strange,...); $\sigma =\uparrow ,\downarrow $ - \textit{spin}. At the
same time the Hamiltonian $\hat{H}_{CS}$ (for instance the Nambu-Lasinio
Hamiltonian), which is an analogue of the BCS one in the $CS$ pairing, must
be invariant under the symmetry group of strong interactions $%
SU(3)_{c}\otimes SU(3)_{f,L}\otimes SU(3)_{f,R}\otimes U(1)$ - see \cite%
{Rischke}.

Let us mention some application of this model in\textit{\ the solid state
physics} such as\textit{\ bi-layered superconductors} in the presence of the
Zeeman effect ($h=\mu _{B}H$)
\begin{equation}
\hat{H}_{Z}=h\sum_{a=1}^{2}\int d^{d}x[\hat{\psi}_{a,\uparrow }^{\dag }(%
\mathbf{x})\hat{\psi}_{a,\uparrow }-\hat{\psi}_{a,\downarrow }^{\dag }(%
\mathbf{x})\hat{\psi}_{a,\downarrow }(\mathbf{x})].  \label{Zeeman}
\end{equation}%
It turns out that the paramagnetic field $h$ can compensate the "isotopic"
mismatch of the chemical potentials ($\delta \neq 0$) giving rise to some
novel effects \cite{Ku-Hof-SSC}. For instance, in case when $\delta >\Delta
_{0}$ superconductivity is destroyed by the isotopic-splitting $\delta $,
while the Zeeman term can compensate this mismatch and thus inducing
superconductivity for $\delta -h<\Delta _{0}$ - the \textit{reentrant
superconductivity }\cite{Ku-Hof-SSC}. Recently, similar effects were found
to exist in layered superconductors placed in the Zeeman field $h$ \cite%
{Buzdin-layered}. In that case the bi-layer superconductivity was studied in
the model (similar to that in Eq.(\ref{H0}-\ref{BCS})) with the \textit{%
intralayer} pairing $\Delta $ only. In that case the interplane hoping ($t$)
in the bi-layer system plays the role of the "isotopic" splitting parameter,
i.e. $\delta =t$, and $\varepsilon _{a=1,2}(\mathbf{\hat{p}})$ are\ the
bonding ($B$) and antibonding ($A$) bands, respectively. It turns out that
for $h<\Delta $ the bi-layer superconductivity is realized, i.e, $\Delta
\neq 0$, while for $h>\Delta $ it is destroyed, i.e., $\Delta =0$. However,
for $h\approx t(\gg \Delta _{0})$ superconductivity appears again (reentrant
superconductivity) on the expense of the $\pi $-phase difference between the
order parameters on the bilayer (in the up- and down-plane), i.e. $\Delta
_{up}=-\Delta _{down}$. In the language of the $ISP$ model there is a
pairing between the antibonding and bonding levels only, i.e., $\Delta
_{AB}\neq 0$ while $\Delta _{AA}=\Delta _{BB}=0$. This result is of a
potential interest in high-temperature superconductors, especially in
bi-layer BISCO.

One expects that similar effects can be realized also in the \textit{quark
matter} with the color superconductivity ($CS$) with two flavor quarks, such
as pulsars-magnetars, where the $CS$ gap is $\Delta \sim 10$ $MeV$ which is
of the order of mass difference $\delta M\sim 5$ $MeV$ between up and down
quark. Having in mind that the magnetic field in magnetars is huge, $%
B>10^{15}G$, then the above compensating effect (by magnetic field and
Zeeman effect) might be operative there.

The $ISP$ model renders a number of interesting effects and some
of them were discussed in \cite{Ku-Hof-SSC}. In the following we
study systems with two levels $a=1,2$ only and apply it to the
ultracold fermionic gases. It is assumed pairing between two
species $\Delta _{ab}\neq 0$ for $a\neq b$, only. This two-state
model is realistic approximation since in diluted gases, for
instance in $^{6}Li$, the interaction of atoms in different
hyperfine states is several order of magnitude larger than in the
same states \cite{Abraham}. Recently, the fermionic superfluidity
was realized recently by several groups \cite{Regal}. However, the
definitive experimental prove of the fermionic superfluidity was
given recently in the remarkable experiment of the Ketterle's
group \cite{Ketterle}. They cooled $^{6}Li$ in magnetic trap below
the Fermi degeneracy. This Fermi cloud, consisting of
approximately$\ 10^{6}$ atoms with two lowest hyperfine states
$\mid 1\rangle $ and $\mid 2\rangle $, was loaded into an optical
dipole trap. Between these states there is a Feshbach resonance at
$B_{0}=875$ $G$. The
BEC-BCS transition occurs in the region between 780 G and 925 G, where for $%
B>$ $B_{0}$ the system is in the BCS side. Since the (attractive) coupling
between atoms in states $\mid 1\rangle $ and $\mid 2\rangle $ at low T is
strong in the $s$-channel, this means that the spatial part of the two-body
wave function $\Psi (a_{1},a_{2})$ is symmetric under the exchange of atoms,
i.e. $\Psi (a_{1},a_{2})=\Psi (a_{2},a_{1})$. As the result the spin part of
the wave function is antisymmetric, i.e.,
\begin{equation}
\mid \left\{ 1,2\right\} \rangle =\frac{1}{\sqrt{2}}[\mid 1\rangle \mid
2\rangle -\mid 1\rangle \mid 2\rangle ].  \label{Spin}
\end{equation}%
The pairing is strong between the states $\mid 1\rangle $ and
$\mid 2\rangle $, i.e., $\Delta _{12}\neq 0$ while $\Delta
_{11}=\Delta _{22}=0$. The crucial observation in \cite{Ketterle}
was the creation of vortices in the rotating systems above some
rotation frequency, what can be considered as a definite prove for
fermionic superfluidity.

Since the $ISP$\ model strongly reminds on metallic superconductors with the
Zeeman (paramagnetic) effect, this analogy might be useful in studying
ultracold fermionic gases. \textit{First}, in the case when the "isotope"
splitting is larger than the critical value $\delta _{c1}=0.71\Delta
_{ab}^{0}$, then in the region $\delta _{c1}<\delta <\delta _{c2}$ (at $T=0$
$K$) a \textit{nonmagnetic analogue} of the $LOFF$ phase is realized with
the nonuniform order parameter $\Delta (\mathbf{x})\sim \cos \mathbf{Qx}$.
For $\delta =\delta _{c1}$ the modulation vector $\mathbf{Q}$ is given by $%
Q_{c1}=2.4(\delta _{c1}/V_{F})$ where $V_{F}$ is the Fermi velocity. It is
known that the structure of the order parameter may contain more wave
vectors $\mathbf{Q}_{i}$ thus making the so called crystalline structure
more favorable \cite{LO-paper} - this (still unsolved) problem is not the
subject of this paper. Concerning the critical value $\delta _{c2}$ it
depends on dimensionality of the system - in 3D one has $\delta
_{c2}=0.755\Delta _{ab}^{0}$, in 2D $\delta _{c2}=\Delta _{ab}^{0}$ while
for 1D $\delta _{c2}\rightarrow \infty $. Note, that for $\delta <\delta
_{c1}$ (at $T=0$ K) the system is in uniform superfluid (superconducting)
state, i.e. $\Delta _{ab}=const$. The transition at $\delta _{c2}$ depends
on the type of the nonuniform solution - for instance for the one plane
solution we use here (only for simplicity) the transition at ($T=0$, $\delta
_{c2}$) is second order. The phase diagram in the plane ($T$, $\delta $) of
the "isotopic" $LOFF$\ phase is shown in $Fig.1$.

\begin{figure}[tbp]
\resizebox{0.6 \textwidth}{!} {
\includegraphics*[width=2cm]{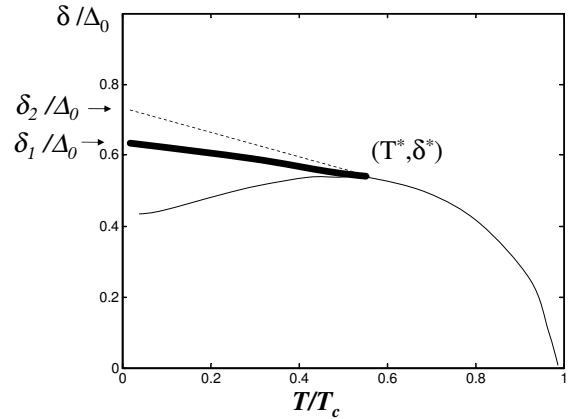}}
\caption{The phase diagram ($\protect\delta, T$) of the 3D superfluid
ultracold fermionic gas. At $T^{\ast }$, $\protect\delta ^{\ast }$ there is
second order phase transition. The bold line is the first order line between
normal state and uniform superconductivity. The region between the
upper-dotted and the lower bold line is the nonuniform LOFF state.}
\label{Fig1-LOFF}
\end{figure}

Second, it turns out that the $LOFF$ state appears at temperatures $%
T<T^{\ast }$ and in the phase diagram there is a tricritical Lifshitz point (%
$T^{\ast }=0.56T_{c}$ , $\delta ^{\ast }=1.32T_{c}$). Near this point\ one
can perform the Ginzburg-Landau ($GL$) expansion analogously to that
obtained in \cite{BuKu-GLexpansion}, \cite{Ku-Hof-SSC}. In order to prove
the existence of the $\pi $- junction in the superfluid $UCFAG$ we shall
study the nonuniform superfluidity near the point ($T^{\ast }$, $\delta
^{\ast }$) where it is straightforward to derive the $GL$ expansion by using
Eqs.(\ref{BCS}-\ref{gap}). As the result the (dimensionless) density of the
free-energy functional $\tilde{F}\{\psi (\mathbf{x})\}\equiv F/E_{c}$ (
where $E_{c}=N(0)\Delta _{0}^{2}/2$ is the condensation energy and $\psi (%
\mathbf{x})=\Delta (\mathbf{x})/\Delta _{0}$) is given by \cite%
{BuKu-GLexpansion}, \cite{Ku-Hof-SSC}
\[
\tilde{F}\{\psi (\mathbf{x})\}=\tau \mid \psi \mid ^{2}-\frac{\gamma }{2}%
\mid \psi \mid ^{4}+\frac{\nu }{3}\mid \psi \mid ^{6}
\]%
\begin{equation}
-\alpha \xi _{0}^{2}\mid \nabla \psi \mid ^{2}+\beta \xi _{0}^{4}\mid \nabla
^{2}\psi \mid ^{2}+\frac{\mu }{2}\xi _{0}^{2}\mid \psi \mid ^{2}\mid \nabla
\psi \mid ^{2}+...  \label{F}
\end{equation}%
Here, $\tau =[T-T_{c}(\delta )]/T^{\ast }$, $\gamma =\tilde{\gamma}(\frac{%
\delta -\delta ^{\ast }}{\delta ^{\ast }}-\gamma ^{\prime }\tau )$, $\alpha =%
\tilde{\alpha}(\frac{\delta -\delta ^{\ast }}{\delta ^{\ast }}-\alpha
^{\prime }\tau )$, where $T_{c}(\delta )$ is the transition temperature from
the normal ($N$) to the uniform BCS state but in the absence of the $LOFF$
state. $N(0)$ and $\xi _{0}$ are the density of states and superconducting
(superfluid) coherence length (at \ $T=0$), respectively. The parameters $%
\tilde{\alpha}$, $\alpha ^{\prime }$, $\tilde{\gamma}$, $\gamma ^{\prime }$,
$\nu $ and $\mu $ are of the order of one and their precise values can be
found in \cite{Buzdin-Houzet}. Since the term in front of $\mid \nabla \psi
\mid ^{2}$is negative it leads to the nonuniform $LOFF$ state near the
tricritical point ($T^{\ast }$; $\delta ^{\ast }$). The minimization of $%
F\{\psi (\mathbf{x})\}$ gives the transition temperature $T_{c,q}(\delta )$
from the normal to the $LOF$F state (the dotted line in $Fig.1$) and the
wave vector $Q$
\begin{equation}
\tau (T_{c,q})=\frac{T_{c,q}(\delta )-T^{\ast }}{T^{\ast }}=\frac{\tilde{%
\alpha}^{2}}{4\beta }(\frac{\delta -\delta ^{\ast }}{\delta ^{\ast }})^{2}
\label{Temp}
\end{equation}%
\begin{equation}
Q^{2}=\xi _{0}^{-2}\frac{\tilde{\alpha}}{2\beta }\frac{\delta -\delta ^{\ast
}}{\delta ^{\ast }}.  \label{Q-vec}
\end{equation}%
In the $LOFF$ state the mass-current density ($\mathbf{j}^{(m)}\equiv
\mathbf{j}/(-2e)$ and $\mathbf{\tilde{j}}^{(m)}=\mathbf{j}^{(m)}/E_{c}$) has
much reacher form then the standard G-L expression

\[
\mathbf{\tilde{j}}^{(m)}=i\mathbf{[}-\alpha \xi _{0}^{2}+\frac{\mu }{2}\xi
_{0}^{2}\mid \psi \mid ^{2}\mathbf{]}\psi ^{\ast }\nabla \psi
\]%
\begin{equation}
+\beta \xi _{0}^{4}i\{\nabla \lbrack \psi \nabla ^{2}\psi ^{\ast }]-2\nabla
\psi \nabla ^{2}\psi ^{\ast }\}+c.c.  \label{J}
\end{equation}%
In the case of charged superconductors the orbital effect of magnetic field
can be accounted for by replacing $\nabla \rightarrow \nabla +(2ie/c)\mathbf{%
A}$. In the case of $UCFAG$, such as $^{40}K$ and $^{6}Li$, the current $%
\mathbf{j}^{(m)}$ is the \textit{mass-current}.

\section{Josephson effect in ultracold fermionic gases}

We shall demonstrate below that in principle it is possible to have a $\pi $-%
\textit{Josephson contact (weak link) }based on cold fermionic gases which
is analogous to the so called $SFS$ contacts (weak links) in the solid state
physics, where in the ferromagnetic weak link ($F$) electrons with different
spin projections have different energies (Zeeman-mismatch). In this case the
superconducting amplitude $F_{\uparrow \downarrow }$ is induced in $F$ by
the proximity effect. This amplitude not only decays (like in the case of
normal metal $SNS$ contacts) but also oscillates between two superconducting
banks, i.e. $F_{\uparrow \downarrow }(x)\sim e^{-Q_{1}x}\cos Q_{2}x$ - see
\cite{Bu-Bu-Pa}, \cite{Buzdin-Review}. The oscillation in the weak link,
especially in the case $\delta \gg \Delta $, is in fact a reminiscence of
the $LOFF$ state as it was shown in \cite{BuKu-GLexpansion} by using Eqs. (%
\ref{F}-\ref{J}).

The Josephson weak-link with $UCFAG$ can be in principle realized by using
various optical and magnetic trap techniques. If the $UCFAG$ with only two
hyperfine levels are trapped their pairing interaction can be described in
some parameter range by Eqs.(\ref{BCS}-\ref{gap}). We further assume that
the left and right part of the weak-link are similar and that the mismatch
of the chemical potentials in them is small, $\delta \ll \Delta
_{0,L}=\Delta _{0,R}$. In that case the left ($x<-L$) and right ($x>L$) BCS\
superfluid-condensates are uniform with the critical temperature $T_{s}$ and
the order parameters $\mid \Delta \mid e^{i\varphi _{L}}$ and $\mid \Delta
\mid e^{i\varphi _{R}}$, respectively. Furthermore we assume that the
weak-link $M$ ($-L<x<L$) with the width $2L$ between the left and right
banks is made of $UCFAG$ with the mismatch $\delta \neq 0$ but which is in
the normal state. In the following we call such a weak-link $SMS$, where $S$
means the BCS superfluid and $M$ is the weak link with the mismatch
parameter $\delta $ and at temperatures where the system is a \textit{normal
Fermi gas}. This means that in $M$ the order parameter $\Delta _{M}=0$ but
the anomalous Green's function $F_{ab,M}(x)\neq 0$ due to the proximity
effect \cite{Buzdin-Review}. At the end we shall argue that the Josephson
effect can exist even in the case $\delta \gg \Delta $, i.e. far away of the
$LOFF$ state. However, in this case one should use rather sophisticated
microscopic many-body techniques - for instance the Eilenberger
quasiclassical equations. In order to prove the existence of the $\pi $%
-Josephson junction we shall study the problem when $T$ and $\delta $ are
just near the tricritical point ($T^{\ast }=0.55T_{c}$ ; $\delta ^{\ast
}=1.32T_{c}$) - see $Fig.1$. In such a case the weak link $M$, which is in
the normal state but very near to the LOFF state, is characterized by the
induced order parameter $\psi _{M}$ which is small near the point ($T^{\ast
} $; $\delta ^{\ast }$) and the $GL$ equation can be linearized. Since $\psi
_{M}(x)$ changes along the $x$-axis which is perpendicular to the surface of
the weak link ($M$) - see $Fig.2$, the $GL$ equation in the weak link $M$
reads
\begin{equation}
\tau \psi _{M}-\alpha \xi _{0}^{2}\frac{\partial ^{2}\psi _{M}}{\partial
x^{2}}+\beta \xi _{0}^{4}\frac{\partial ^{4}\psi _{M}}{\partial x^{4}}=0.
\label{WL}
\end{equation}%
In order to simplify the analysis we assume that $\tau \gg \alpha ^{2}/\beta
$ and the solution $\psi _{M}\sim e^{Q_{M}x}$ oscillates with $Q_{M}=\xi
_{0}^{-1}(1\pm i)(\tau /4\beta )^{1/4}$. The general solution in the weak
link is%
\begin{equation}
\psi _{M}(x)=\sum_{p=1,-1}(A_{p}e^{pQ_{M}x}+B_{p}e^{pQ_{M}^{\ast }x})
\label{PsiM}
\end{equation}%
where $A_{p}$ and $B_{p}$ can be obtained from the boundary conditions at $%
x=-L$ and $x=L$. Since at present there is no microscopic theory for the
Josephson effect in $UCFAG$ we use the experience from the physics of the $%
SFS$ weak links. There various boundary conditions do not destroy the
oscillations of $\psi (x)$ in the weak link, which is important property for
the realization of the $\pi $-contact. Therefore, we assume that $\psi
_{M}(x)$ and $\partial \psi _{M}/\partial x$ are continuous on boundaries at
$-L$ and $L$.

\begin{figure}[tbp]
\resizebox{0.5 \textwidth}{!} {
\includegraphics*[width=3cm]{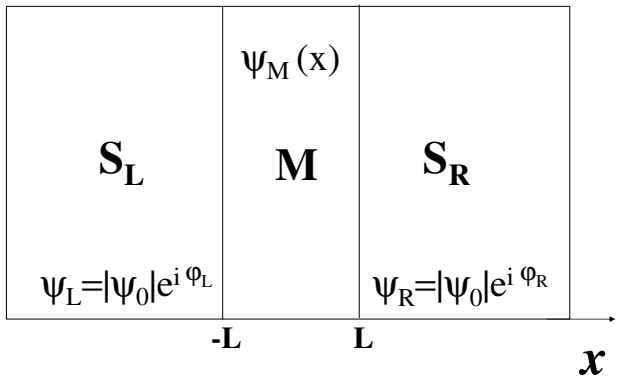}}
\caption{The Josephson junction with uniform superfluidity in banks $S_{L,R}$
with $\mid\protect\psi _{L,R}\mid=const$ and $\protect\delta <\protect\delta %
^{\ast } $. In the weak link $M$ with the thickness $2L$, which is in the
normal state (above the doted line in $Fig.1$), one has $\protect\delta \neq
0$ and $\Psi _{M}$ oscillates and decays. The Josephson current flows along
the $x$-axis.}
\label{Fig2-JJ}
\end{figure}

Since we study the problem near the tricritcal point ($T^{\ast }$, $\delta
^{\ast }$) then the standard term ($\sim \psi _{M}\nabla \psi _{M}^{\ast }$)
in the current is small and the linearized expression for the current $%
\tilde{j}_{x}^{(m)}$ is given by%
\begin{equation}
\tilde{j}_{x}^{(m)}\approx i\beta \xi _{0}^{4}[\psi _{M}\frac{\partial
^{3}\psi _{M}^{\ast }}{\partial x^{3}}-\frac{\partial \psi _{M}}{\partial x}%
\frac{\partial ^{2}\psi _{M}^{\ast }}{\partial x^{2}}+c.c].  \label{Jx}
\end{equation}%
Due to the particle current conservation the Josephson current trough the
weak link $M$ can be calculated at the midpoint $x=0$, i.e., $\tilde{j}%
^{(m)}(\varphi )=\tilde{j}_{x}(x=0)$ is given by%
\begin{equation}
\tilde{j}^{(m)}(\varphi )=\tilde{j}_{c}\sin \varphi ,  \label{J-fi}
\end{equation}%
where%
\[
\tilde{j}_{c}=4\mid Q_{M}\mid ^{3}\beta \xi _{0}^{4}\mid \psi _{0}\mid
^{2}e^{-\sqrt{2}\mid Q_{M}\mid L}
\]%
\begin{equation}
\times \sin (\sqrt{2}\mid Q_{M}\mid L-\frac{\pi }{4}).  \label{Jc}
\end{equation}%
Here, $\varphi =\varphi _{L}-\varphi _{R}$ is the phase difference on th
junction. From Eq. (\ref{Jc}) one concludes that $\tilde{j}_{c}$ can reach
negative values whenever $\sin (\sqrt{2}\mid Q_{M}\mid L-\frac{\pi }{4})<0$.
This can be achieved by changing either $T$, $\delta $ or the thickness $L$
of the weak link $M$. In such a way one can realize a $\pi $\textit{%
-junction }- the contact with $\tilde{j}_{c}<0$. To remaind the reader, the $%
\pi $-contact is characterized by the phase difference $\varphi =\pi $ in
the ground state, while in the standard Josephson contacts $\varphi =0$
minimizes the energy of the contact. In the next Section we are going to
show, that if the $\pi $-junction is placed in a ring with the large radius $%
R$ there is a \textit{spontaneous superfluid flow} through the ring, i.e. the%
\textit{\ time-reversal symmetry is broken spontaneously}.

\section{$\protect\pi $-junction and spontaneous superfluid flow}

Let us consider a loop made of an ultracold fermionic BCS superfluid with
the $\pi $-contact placed in a ring of radius $R$ as shown in $Fig.3$.

\begin{figure}[tbp]
\resizebox{0.5 \textwidth}{!} {
\includegraphics*[width=2cm]{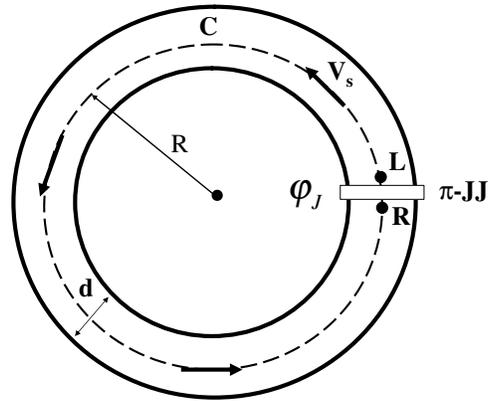}}
\caption{The ring with the $\protect\pi $-Josephson junction. $\protect\phi %
_{J}$ is the phase on the contact. The closed contour is $C$ and for $d<<R$
the superfluid velocity $V_{s}$ is uniform over the cross-section of the
ring. For the radius $R>R_{c}$ a spontaneous current flow with the velocity $%
V_{s}$ - the spontaneous breaking time-reversal symmetry.}
\label{Fig3-pi-ring}
\end{figure}

If the mass-current flows through the Josephson junction and the ring the
total energy (per cross-section $S$ of the ring) of the system $%
W(=W_{K}+W_{J})$ \ is due to the kinetic energy $W_{K}=E_{K}/S$ of
the \textit{circulating superfluid current} and the energy
$W_{J}=E_{J}/S$ of the Josephson contact. In the ring with a small
cross-section ( $d\ll R$) the superfluid velocity $V_{s}(r)$ is
practically constant over the cross-section of the ring
($V_{s}(r)\approx V_{s}(R)\equiv V_{s}$) and the total energy is
given by the simple expression
\begin{equation}
W=\frac{n_{s}(2m_{a})V_{s}^{2}}{2}(2\pi R)-\frac{\hslash \mid
j_{c}^{(m)}\mid }{2}cos(\varphi _{J}+\varphi _{int}).  \label{W}
\end{equation}%
Here, $n_{s}$ is the density of superfluid pairs, $m_{a}$ is the mass of the
fermionic atom, and $\varphi _{J}=\varphi _{L}-\varphi _{R}$ is the phase
difference on the contact. The intrinsic phase of the junction is $\varphi
_{int}=0$ for $j_{c}^{(m)}>0$ and $\varphi _{int}=\pi $ for $j_{c}^{(m)}<0$.

The relation between the superfluid velocity $\mathbf{V}_{s}=(\hslash
/2m_{a})\nabla \varphi $ and the phase $\varphi _{J}$ can be obtained from
the \textit{quantization condition} of the superfluid phase $\varphi $ along
the closed path $C$ in the ring - see $Fig.3$%
\begin{equation}
2\pi n=\doint\limits_{C}\delta \varphi =\dint\limits_{C^{\prime }}\nabla
\varphi d\mathbf{l}+\varphi _{J}.  \label{Cont}
\end{equation}%
The path $C^{\prime }$ goes from $L$ to $R$ - see $Fig.3$. As the result one
obtains the relation between $V_{s}$ and $\varphi $
\begin{equation}
2\pi \frac{V_{s}}{V_{R}}=-\varphi _{J}+2\pi n,  \label{Phase}
\end{equation}%
where $V_{R}=\hslash /2m_{a}R$. The dimensionless energy of the junction
(contact) $\bar{W}=W/\pi \hslash n_{s}V_{R}$ is given by%
\begin{equation}
\bar{W}=\left( \frac{V_{s}}{V_{R}}\right) ^{2}-\frac{\mid j_{c}^{(m)}\mid }{%
2\pi n_{s}V_{R}}\cos (2\pi \frac{V_{s}}{V_{R}}+\varphi _{int}).  \label{Ener}
\end{equation}%
The minimization of $\bar{W}$ with respect to $V_{s}$ gives the condition on
the ring radius $R$ for the appearance of a spontaneous superflow with $%
V_{s}\neq 0$
\begin{equation}
R>R_{c}=\frac{\hslash }{\pi 2m_{a}V_{J}}.  \label{Rc}
\end{equation}%
Here, $V_{J}=j_{c}^{(m)}/n_{s}$ is an effective velocity trough the weak
link. In that case the $\pi $-contact ($\varphi _{int}=\pi $) is realized,
i.e. there is a \textit{spontaneous superfluid flow in the ring}. This also
means that there is a \textit{spontaneous breaking of thee time-reversal
symmetry} in the system. From Eq.(\ref{Rc}) one concludes that for small
critical currents (velocity $V_{J}$) of the weak link the $\pi $-contact is
realized for large $R$. Due to the lack of precise experimental data for
parameters $V_{F}$, $V_{J}$, $\Delta $, $E_{F}$ one can make some
qualitative guesses of $R_{c}$ only. For instance in $UCFAG$ with $^{40}K$,
by assuming that $E_{F}\sim (1-10)$ $\mu K$, $(\Delta /E_{F})\sim
10^{-2}-10^{-3}$, and that $V_{J}\sim (10^{-2}-10^{-4})V_{s,c}$ with $%
V_{s,c}\approx (\Delta /E_{F})V_{F}$ - the critical depairing velocity, one
gets $R_{c}\sim (10^{3}-10^{5})$ $\mu m$. Since $R_{c}$ scales inversely
with the atomic mass it is larger for lower atomic mass.

\section{Discussion and conclusions}

In conclusion, we have shown that in ultracold fermionic atomic gases ($%
UCFAG $) with the mismatch in energies (chemical potentials) of pairing
atoms one can realize $\pi $-Josephson junction (contact) in $SMS$ weak
links. In that case the weak link $M$ is made from $UCFAG$ with the finite
mismatch in hyperfine energies (chemical potentials) of the pairing atoms, $%
\delta \neq 0 $. Our proof was given for the case when the energy mismatch $%
\delta $ is of the order of the fermionic superfluid gap $\Delta $, $\delta
\sim \Delta $, in which case $M$ is in the normal state but near to the
nonuniform $LOFF$ state. However, by analogy with the physics of $SFS$
junctions one expects that the $\pi $-Josephson junction can be realized
also for very large mismatch, i.e., $\delta \gg \Delta $. In this case the
decay-length of the superfluid pairing amplitude induced in the weak link $M$
is very short, i.e. $Q_{M}^{-1}\sim (V_{F}/\delta )\ll \xi _{0}$. In such a
case the problem must be attacked by more sophisticated theoretical methods,
for instance by the Eilenberger quasiclassical equations in case when $%
q^{-1}\gg k_{F}^{-1}$.

Concerning a possible realization of the $\pi $-contact in $UCFAG$ it seems
that traps by optical lattices are more promising way than inhomogeneous
magnetic traps. The possibility for the realization of $\pi $-junctions in $%
UCFAG$ opens a room for numerous speculations about potential applications.
For instance, various contacts of the type $SM_{1}M_{2}S$ as well as
trilayer systems such as $M_{1}SM_{2}$would allow interesting switching
properties of these systems. These systems, if they are realizable, would
offer a possibility for a new kind of electronics - \textit{hypertronics}.
For instance, the computer memory might be realized in weak links $M_{1,2}$,
while the logic processing can be done by manipulating the superfluid state.
These are open but very attractive problems in this very promising field.
One should admire that the realization and manipulation of Josephson
junctions in $UCFAG$ represents a considerable experimental challenge.

\textbf{Acknowledgement}. I express my deep gratitude to I. Bo\v{z}ovi\'{c},
A. I. Buzdin, P. Fulde, W. Hofstetter, I. Kuli\'{c}, D. Pavuna and D.
Rischke for support. I am thankful to Cristian Schunk for critical comments
concerning the experimental confirmation of the fermionic superfluidity in $%
^{6}Li$.

\end{document}